# Metamorphosis in Carbon Network: From Penta-Graphene to Biphenylene under Uniaxial Tension

Obaidur Rahaman*[,1] Bohayra Mortazavi[*,1], Arezoo Dianat[2], Gianaurelio Cuniberti[2], and  Timon Rabczuk[*,3]

[1]*Institute of Structural Mechanics, Bauhaus-Universität Weimar, Marienstr. 15, D-99423 Weimar, Germany*

[2]*Institute for Materials Science and Max Bergmann Center of Biomaterials, TU Dresden, D-01062 Dresden, Germany*

[3]*College of Civil Engineering, Department of Geotechnical Engineering, Tongji University*

*Obaidur Rahaman, Email: ramieor@gmail.com

**Abstract**

The power of polymorphism in carbon is vividly manifested by the numerous applications of carbon-based nano-materials. Ranging from environmental issues to biomedical applications, it has the potential to address many of today's dire problems. However, an understanding of the mechanism of transformation between carbon allotropes at a microscopic level is crucial for its development into highly desirable materials. In this work we report such a phase transformation between two carbon allotropes, from penta-graphene (a semiconductor) into biphenylene (a metal) under uniaxial loading. Using density functional theory we demonstrated that the phase transformation occurs through a synchronized reorganization of the carbon atoms with a simultaneous drop in energy. The results of this work confirms that penta-graphene is a meta-stable structure. On the other hand, a rigorous analysis of biphenylene suggests that it is an energetically, mechanically, dynamically and thermally stable structure, both in the form of a sheet and a tube. Its electronic structure suggests that it is metallic in both these forms. Therefore, this work unravels the possibility of phase transition in 2-D carbon systems and thereby designing nano-materials capable of altering their properties in an instant. Furthermore, heating biphenylene sheet at a high temperature (5000K) revealed another phase transformation into a more stable hexa-graphene like structure. This proposes thermal annealing as a possible method of synthesizing one 2-D carbon allotrope from another.

**Introduction**

The existence of allotropes in nature provides a fantastic opportunity for synthesizing alternative structures of desirable functions. Enabled by the versatile nature of carbon bonding, a rapidly growing number of experimentally isolated and theoretically predicted carbon allotropes are proposed as nano-materials suitable for interesting and unique applications. Graphene,[1] phagraphene,[2] penta-graphene,[3] amorphized graphene,[4, 5] graphenylene[6-8] and Haeckelites[9-11] are among many of the proposed 2D carbon allotropes with notable electronic and mechanical properties.[12-17] Graphene, which is successfully synthesized in year 2004, is the strongest and most stable among all of them.[1, 18]

Although theoretical studies are valuable in understanding the structure-function relationships of these materials at a fundamental level, the experimental synthesis of them still remain a challenge. The experimental synthesis of a predicted structure depends on many factors, specially its stability. Stability is a prerequisite for its formation as well as successful isolation from alternative possible structures. The experimental feasibility not only requires the structure to be at a local minimum on the potential energy surface but its neighboring minima corresponding to alternative structures should also be well separated by high energy barriers.[19] Moreover, the relative thermodynamic stabilities and phases of different allotropes of carbons remain mysterious.[20]

Thus, understanding the relationship and possible transitions between the allotropes of a targeted material is essential for its experimental realization. Recently Zhang et al. predicted the existence of penta-graphene, a metastable new non-planar allotrope of carbon entirely consisting of pentagons.[3] It possesses both $SP^2$ and $SP^3$ hybridization, an unusual negative Poisson's raio and an intrinsic band gap as large as 3.25 eV. They demonstrated the structure to be dynamically and mechanically stable and it could withstand a high temperature of 1000K indicating that the penta-graphene phase was separated from other local minima on the potential energy surface by high energy barrier. The superior and remarkable properties of penta-graphene make it an attractive target of research in the field of nanoelectronics as evident from the numerous recent studies devoted on this allotrope. [21-31]

Considering the highly desirable properties of this theoretically proposed carbon allotrope, an in-depth investigation of its metastable nature and the feasibility of its experimental synthesis is of supreme value. A recent computational study argued that the experimental synthesis of penta-graphene is rather improbable because of energetic funneling toward lower energy graphene structure with entirely $SP^2$ hybridization.[19] In fact, another molecular dynamics study demonstrated that upon uniaxial loading penta-graphene transforms into graphene through stable intermediates consisting of mixtures of pentagons and hexagons.[32] As opposed to this previous study, we report a rather sudden transformation of penta-graphene under uniaxial strain into what is identified as biphenylene.[33-40] The study was conducted using Density Functional Theory. Instead of the gradual and local transformation proposed by the force field study,[32] the phase transition was found to be a global phenomenon where all atoms moved in unison.

The final product biphenylene not only has potential applications in various fields as stated bellow but it is also experimentally synthesized. Different oligomers of biphenylene were synthesized in the lab and characterized as synthetic targets for materials with interesting electronic properties.[39] Recently, an octafunctionalization method has been developed to fabricate and extend biphenylene structures into nanoribbons and sheets.[36] Biphenylene is predicted to be metallic in the form of a sheet or tube[38] but possessing a band gap in the form of ribbon[38] or by functionalization.[35] Biphenylene sheet can be used for hydrogen storage and in Lithium Ion Batteries (LIB) as it is predicted to adsorb lithium significantly stronger than graphene and graphyne.[34] A recent study suggested potential application of biphenylene-based nanoribbons in electronics and optoelectronics in the visible range.[33] Thus, in this work, a thorough characterization of the properties of biphenylene has also been carried out in order to explore its potential applications as a carbon nano-material.

Similar to the transformation observed in the 3D carbon system,[41, 42] this work demonstrates the possibility of phase transition in 2D carbon systems. It not only demonstrates possible pathways of synthesis for predicted 2D-carbon allotropes but also unfolds a novel paradigm in nanomaterials where the properties can be manipulated by phase-transition.

**Methods**

*Vienna ab Initio Simulation Package* (VASP)[43, 44] was used to perform the *ab initio* and AIMD simulations within the framework of DFT. The exchange-correlation energy was treated within the Generalized Gradient Approximation (GGA)[45] using Perdew, Burke, and Ernzerhof (PBE) functional and Projector Augmented Wave (PAW)[46] method was used to calculate the interaction between the frozen core and valence electron. The 2D carbon network was separated from its periodic image by a vacuum of length 20Å in the perpendicular direction. Energy minimizations were performed using conjugate gradient method with a threshold for total energy of $10^{-5}$ eV and atomic force components of 0.02 eV/ Å . An energy cutoff of 450 eV and a k-point mesh size of $7 \times 7 \times 1$ in the Monkhorst-Pack scheme[47] were used for the Brilloin zone sampling. A higher k-point sampling of $30 \times 30 \times 1$ was used for the band structure and density of state calculations. For the AIMD simulations, a time step of 1fs was implemented and Langevin thermostat[48] was used for maintaining the temperature. Phonopy[49] was used for the calculation of phonon properties. Density Functional Perturbation Theory (DFPT)[50] as implemented in VASP and a high-accuracy energy convergence criterion of $10^{-8}$ eV was used for the calculation of atomic forces.

## Results and Discussions

### Penta-graphene transforms into biphenylene under uniaxial strain

Penta-graphene under uniaxial strain is observed to undergo a sudden and global transformation into another allotrope of carbon. A supercell structure of penta-graphene consisting of 72 carbon atoms was constructed and minimized using VASP. Then the structure was treated with a gradual uniaxial strain applied step by step. At each step of the DFT calculation, the strain was incremented by 0.25% followed by a energy minimization with fixed size of the cell. During this process, the energy was observed to gradually rise until about a strain of 20%. At this point we observed a sudden drop in energy along with a global reorientation of the structure into an alternative allotrope of carbon composed of 4,6 and 8-membered carbon rings (Figure 1).

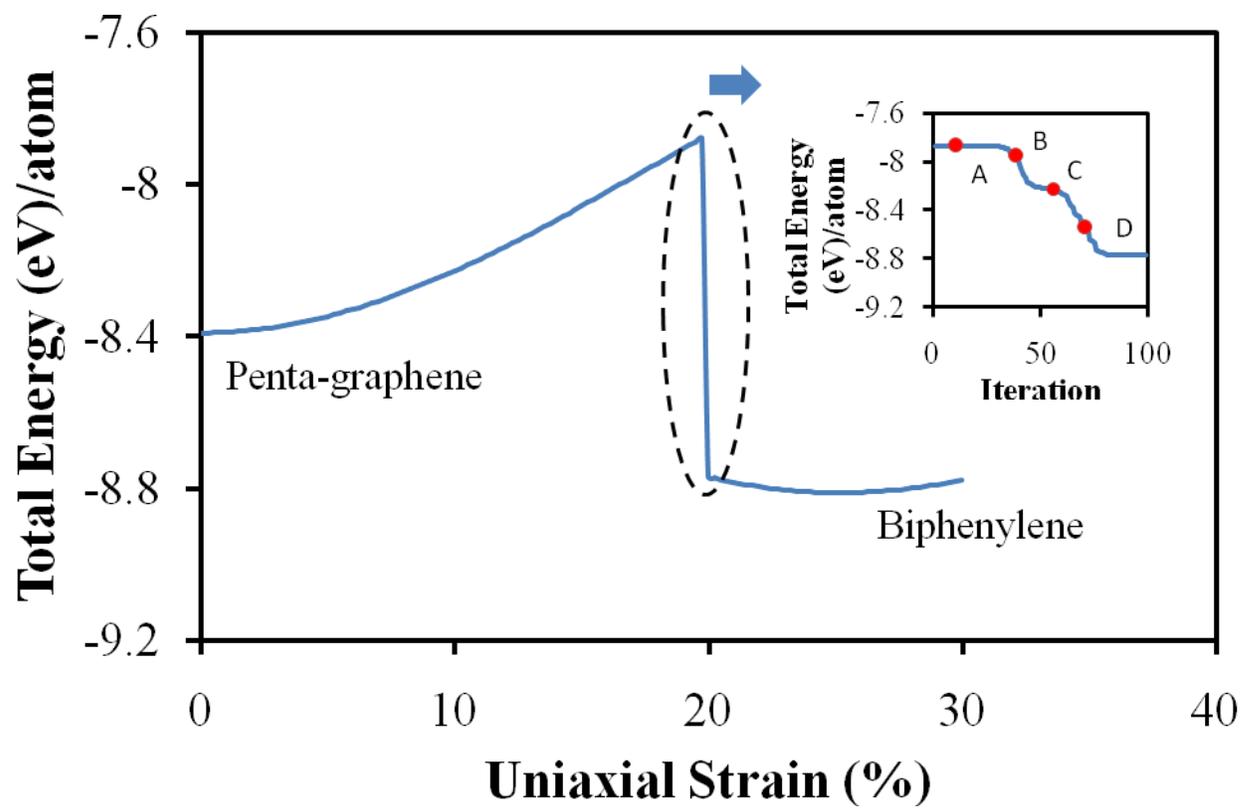

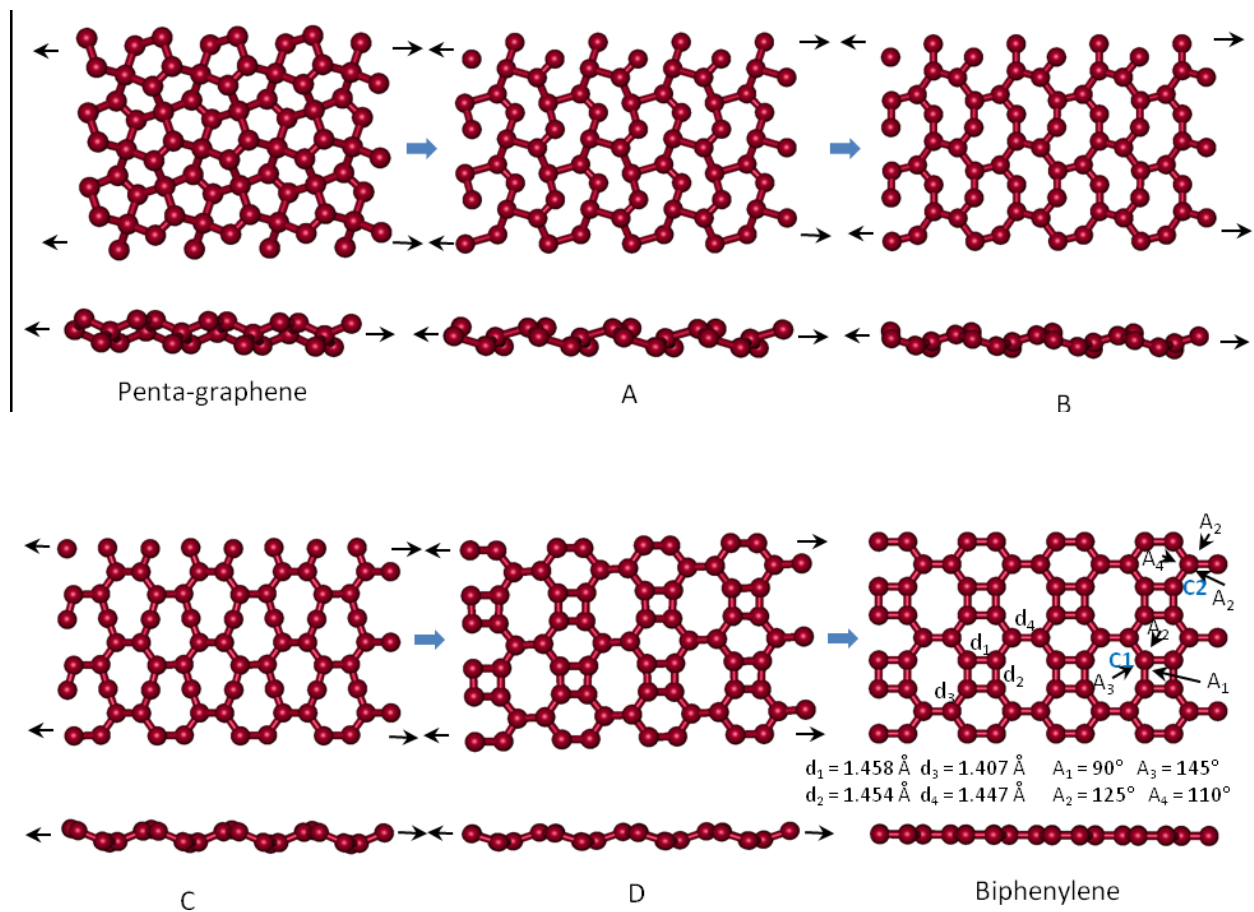

Figure 1: The evolution of energy as penta-graphene is treated with uniaxial strain. The disintegration of penta-graphene with a simultaneous formation of biphenylene can be visualized using the top and side views of the selected intermediates. These intermediates were formed during the minimization at a strain of 20%.

The penta to biphenylene transformation that occurred within a single energy minimization step at 20% strain can be followed from Figure 1 inset and a few intermediate snapshots, A, B, C and D. There were no distinct characteristics in these intermediates. They were selected only for the purpose of demonstration. One structure smoothly transformed into another during the energy minimization step. The penta to biphenylene transformation occurred as a global phenomenon rather than a local one as proposed by the force field study.[32]

We note that Penta-graphene has a non-planar structure with both $sp^2$ and $sp^3$-hybridized carbon atoms (see the side view). During the penta to A transformation, all the $sp^2$-$sp^3$ C-C bonds (but not the $sp^2$-$sp^2$ C-C bonds) that were aligned in the direction of the applied strain dissociated. This created distorted and non planar (see the side view) 8-membered rings in the structure. In the following transformations A→B, B→C and C→D, the 8-membered rings gradually became more symmetric while the structure remained non planar. This was accompanied by a significant drop in energy. The intermediates A, B and C were consisted of only 8-membered rings and nothing else. In the C→D transformation, half of the 8-memberd rings collapsed into a six-membered and a 4-memberd ring by forming a C-C bond. The final step, D→ biphenylene was marked by a non-planar to planar transformation of the structure.

The stress in the structure responded to the increase in the strain in a similar manner as the energy. Before the penta to biphenylene transformation, the increasing uniaxial strain caused a gradual rise in the stress (Figure S1). At strain=20%, when the penta to biphenylene transformation occurred, a sudden drop in stress was observed. The sudden transformation of the non-planar penta-graphene to planar biphenylene structure seems to have caused the negative stress which was relaxed in the subsequent steps.

We note that the strain of 20% is close to the theoretical limit of graphene stretching under uniaxial strain.[51] In another study we compared the stress-strain curves of graphene, pentagraphene, biphenylene and a few other graphene-like carbon and carbon nitride networks under uniaxial strain.[51] Although the maximum stress attained during the stretching was highest for graphene, both pentagraphene and biphenylene demonstrated significant strength. We note that thermal fluctuations in the system can make the attainment of such a high value of strain difficult. In particular uniform distribution of the strain across the system is difficult to achieve in an experimental setup.

The final biphenylene structure was obtained by removing the strains and allowing the system to relax to its low energy local minimum. It is worth noting that the energy of the optimized biphenylene structure is considerably lower than the energy of the optimized penta-graphene structure. Evidently biphenylene is more stable than penta-graphene. It is a planar structure and each carbon atom has a $sp^2$ hybridization. As opposed to the penta-graphene structure, this structure is anisotropic. In its fully relaxed form, the 8-membered rings are elongated in one direction whereas the 6-membered rings are elongated in the other direction. The 4-membered rings are also slightly elongated in the same direction as the six membered rings.

Although biphenylene has 3 types of rings, it has only two types of carbon atoms (Figure 1). Carbon atoms that are part of the 4-memberd rings (C1) and carbon atoms that are not part of the 4-membered rings (C2). Both C1 and C2 have distorted sp2 hybridization. The C-C-C angles around C1 are 90°, 125° and 145° whereas the angles around C2 are 110°, 125° and 125°. The preferred angle of the trigonal planar geometry of sp2 hybridization for minimum electron repulsion is 120°. Thus, we expect more electron repulsion near C1 than C2 since its angles are farther away from this optimum angle of 120°. The maximum electron repulsion is expected at the angle of 90° near C1.

We verified that a transformation to biphenylene also occurred when uniaxial strain was applied to penta-graphene in the perpendicular direction to the one showed in this picture. This was expected since penta-graphene is isotropic in nature. Furthermore, we investigated if a similar transformation would occur if the strain was applied in a biaxial manner. Strains were applied simultaneously in both x and y directions with an increment of 0.25% with subsequent energy minimizations. The calculations were performed until a maximum strain of 30%. Although a non-planar to planer transformation was observed during the evolution of the structure under biaxial strain, no biphenylene structure was obtained as the end product (Figure S2). Thus, it is evident that the pentagraphene to biphenylene transformation occurs only when the strain is applied in a uniaxial manner. The simultaneous relaxation in the perpendicular direction to the applied uniaxial strain is necessary for the evolution of the structure into biphenylene.

Recently, a ReaxFF force field bases molecular dynamics simulation study suggested a gradual penta-graphene to graphene transformation under uniaxial strain.[32] Instead, our DFT study suggests a sudden penta-graphene to biphenylene transformation under uniaxial strain. Although both studies suggest the meta-stable nature of penta-graphene, a non-planar to planar transformation and a release in energy during the process, the difference in the final products can be understood by the difference in the loading conditions applied in these two studies. In the previous ReaxFF force field study thermal fluctuations were allowed while the strain was gradually increased. This could be the reason of the localized penta-graphene to graphene transformation as reported by the study. On the other hand, in the current DFT work the strain was homogenously applied across the system without allowing for the thermal fluctuations. The penta-graphene to biphenylene transformation is a result of the application of such affine stress/strain field. Thus, these two studies demonstrate phase transformations under two different loading conditions with different end results, both valid theoretically and do not contradict with each other.

In the following sections we explore various properties of biphenylene.

**Biphenylene is energetically stable**

The stability of a structure directly correlates with it energy. In general, the lower the energy, the higher is its stability. Figure 2 compares the energy of biphenylene with other 2D carbon allotropes and their dependence on area. [11, 52-56] Only graphene and phagraphene are energetically more favorable than biphenylene. In their fully relaxed states the total energy per atom of graphene, phagraphene and biphenylene are -9.28 eV, -9.08 eV and -8.82 eV, respectively. So biphenylene is only 0.47 eV higher in energy than graphene, the most stable 2-D carbon structure. The other 2D carbon allotropes including penta-graphene and the experimentally observed dodecahedral $C_{20}$ cage are less stable than biphenylene. We note that the carbon allotropes included in this comparison are not exhaustive. However, when compared to another study, biphenylene still belongs to the group of allotropes close to graphene in the energy scale.[18]

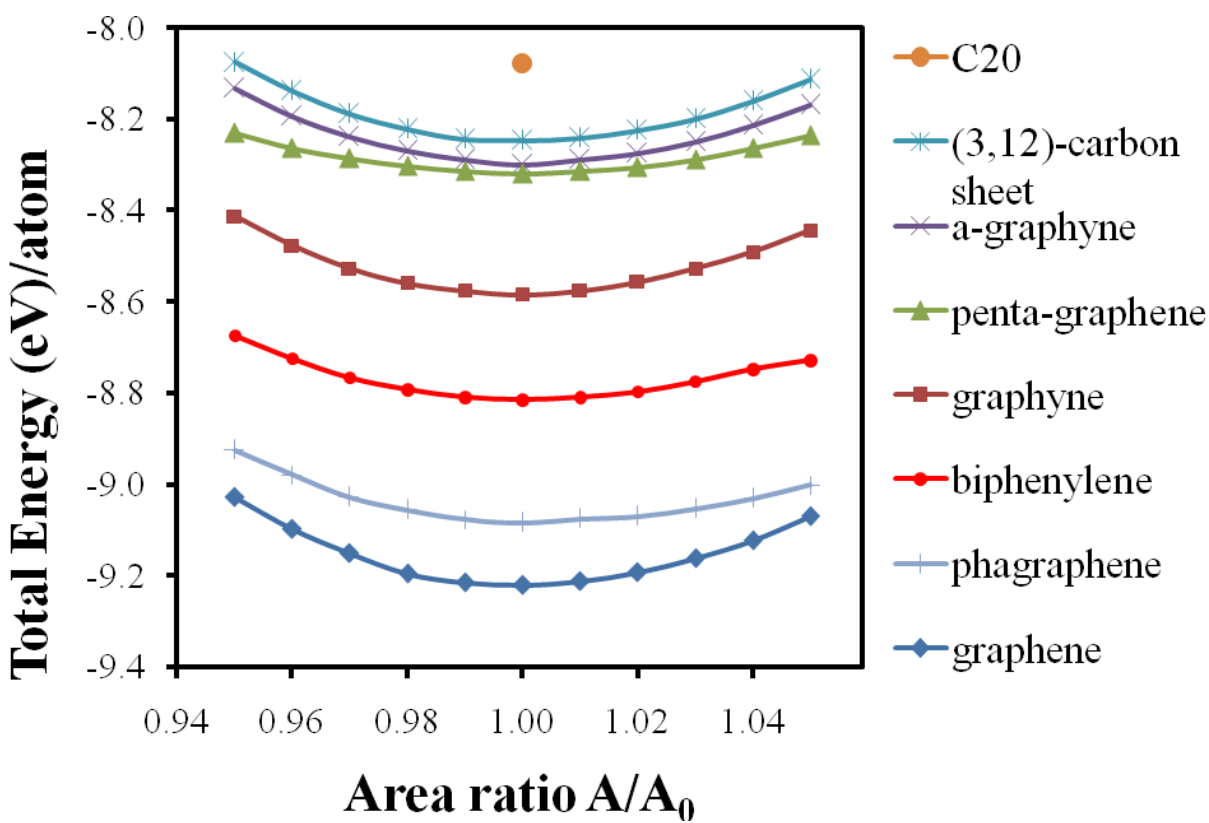

Figure 2: The area dependence of the per atom energies of 2D carbon allotropes. The data for biphenylene and phagraphene were obtained in this work. The rest of the data were taken from reference [3].

So biphenylene is energetically stable which makes its experimental realization possible as confirmed by its successful synthesis by a recently proposed octafunctionalization method.[36]

**Biphenylene is metallic in nature**

In order to understand the electronic properties of biphenylene we calculated its band structure and electronic density of states (DOS). In agreement with previous studies,[35, 38] a high DOS is observed at the Fermi level confirming that biphenylene is metallic in nature. We note that, a DFT study suggested an opening of band gap when biphenylene is constructed as nano-ribbons with armchair like edges (a band gap of 1.71 eV for a width of 6.2 Å).[38]

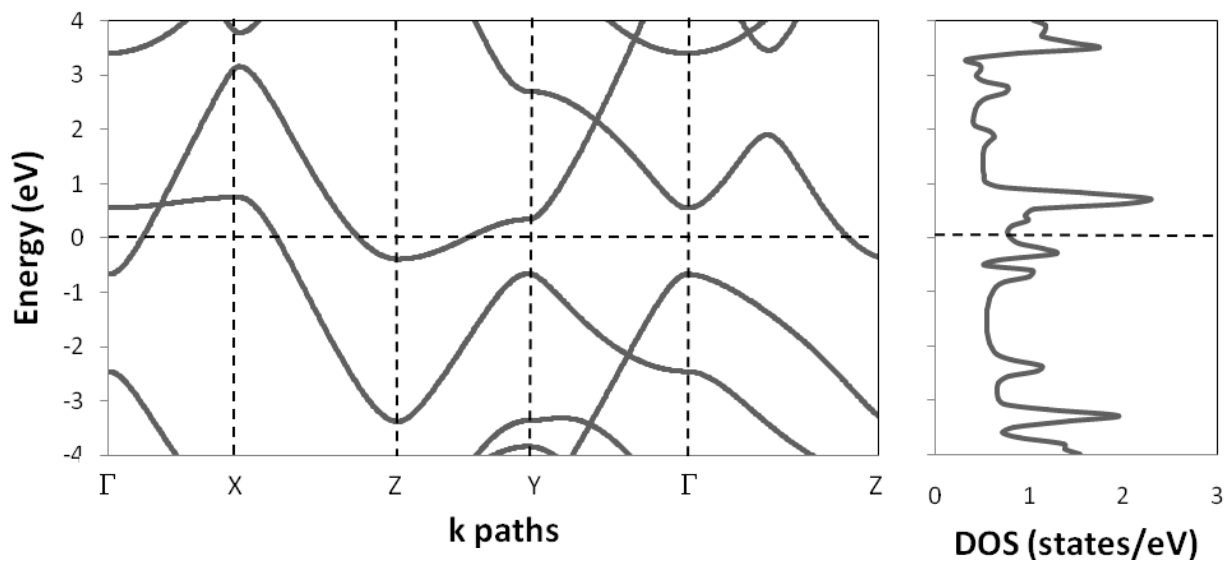

Figure 3: Band structure and density of states (DOS) of biphenylene sheet.

We also note that, although biphenylene is a 2D network of $SP^2$ carbon that exhibits a metallic behavior just like the Haeckelites,[11] it does not belong to this type since it does not have any pentagons and heptagons.

**Biphenylene is dynamically stable**

The lattice dynamics of biphenylene was investigated in order to determine its dynamic stability. Figure 4A shows the phonon dispersion of biphenylene. The absence of any imaginary mode in the entire Brillouin zone suggests that biphenylene is dynamically stable. Three distinct acoustic modes were observed in the phonon spectra of biphenylene. This is similar to the phonon dispersion in graphene.[57, 58] Figure 4B shows the phonon density of state for biphenylene. Phonons of different frequencies ranging from zero to about 50 THz were observed without a gap.

A)

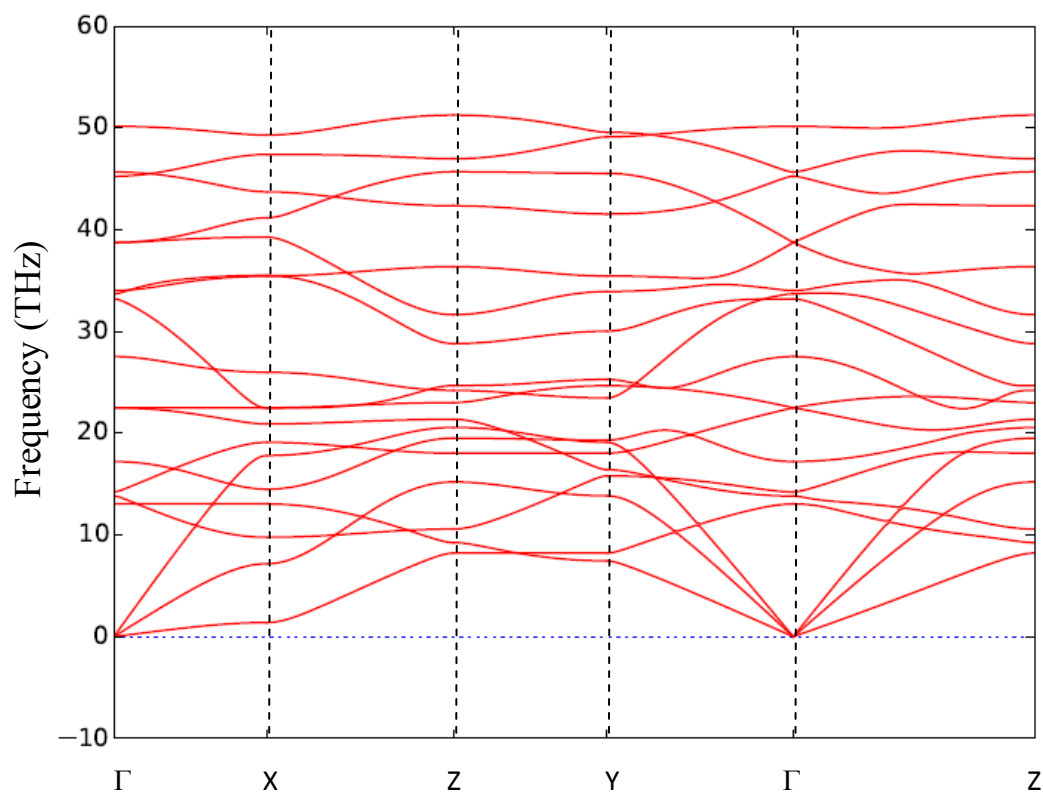

B)

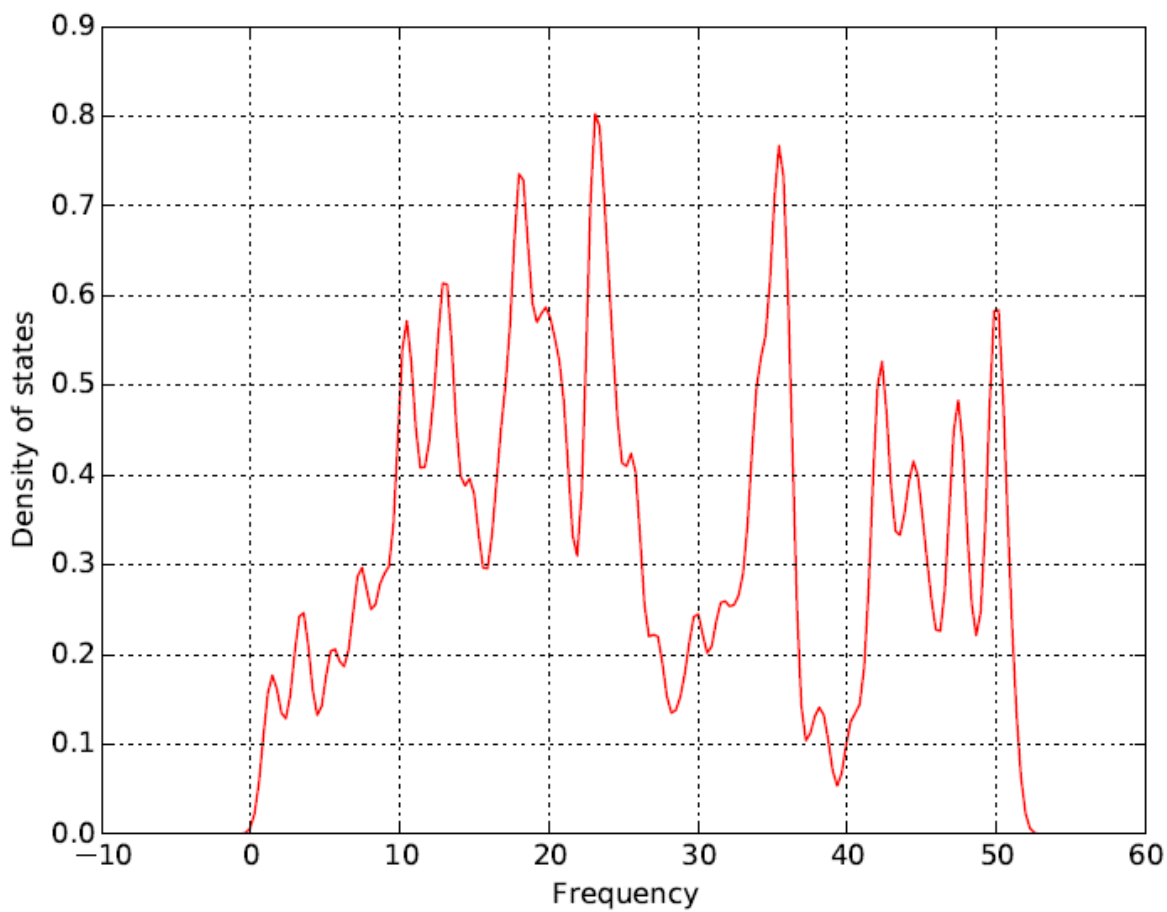

Figure 4: A) Phonon dispersion and B) PhDOS of biphenylene

**Biphenylene is thermally stable but can be transformed at high temperature**

In order to examine the thermal stability of biphenylene, we performed an *ab initio* molecular dynamics (AIMD) simulation starting from the energy minimized structure. The system was heated at room temperature (300K) for 6 ps and with a time step of 1fs. Throughout the simulation the structure remained intact without any structural rearrangements, except slight fluctuations in the 2D surface.

Following this, we examined the stability of biphenylene by running 6ps AIMD simulations at higher temperatures (T=1000K, 2000K etc). At the end of each simulation, the structure remained unchanged even at very high temperature of 4000K. For the last case, the simulation was extended for a total of 12ps. But the structure remained intact. As expected, the 2D surface fluctuations were more pronounced than the one at room temperature. The final snapshots of the AIMD simulations at different temperatures are shown in Figure S3.

The stability of biphenylene at high temperatures suggests that it is separated from its alternative structures in the potential energy surface by high energy barriers. If synthesized successfully, biphenyl will not easily disintegrate into other structures.

We observed a complete disintegration of biphenyl at a higher temperature of 6000K. No regular structure remained in the final product. However, at T=5000K we observed something interesting, a structural transformation of biphenyl instead of complete disintegration (Figure 5). During the simulation multiple bond breakings and formations took place and biphenyl transformed into a 2D-network of carbon with rings of different sizes. This transformation was accompanied by a clear drop in potential energy indicating a stabilization of the structure. A redistribution of ring sizes was observed during the 12 ps of simulation (see the inset of Figure 5) with the 4,6 and 8 membered rings present in the initial structure of biphenylene to a range of ring sizes from 4 to 9. Although 5, 7 and 9 membered rings appeared as new ring sizes, the overall change can be summarized as a narrowing of the distribution around ring size 6, the only one present in the most stable structure of 2D-carbon, hexa-graphene.

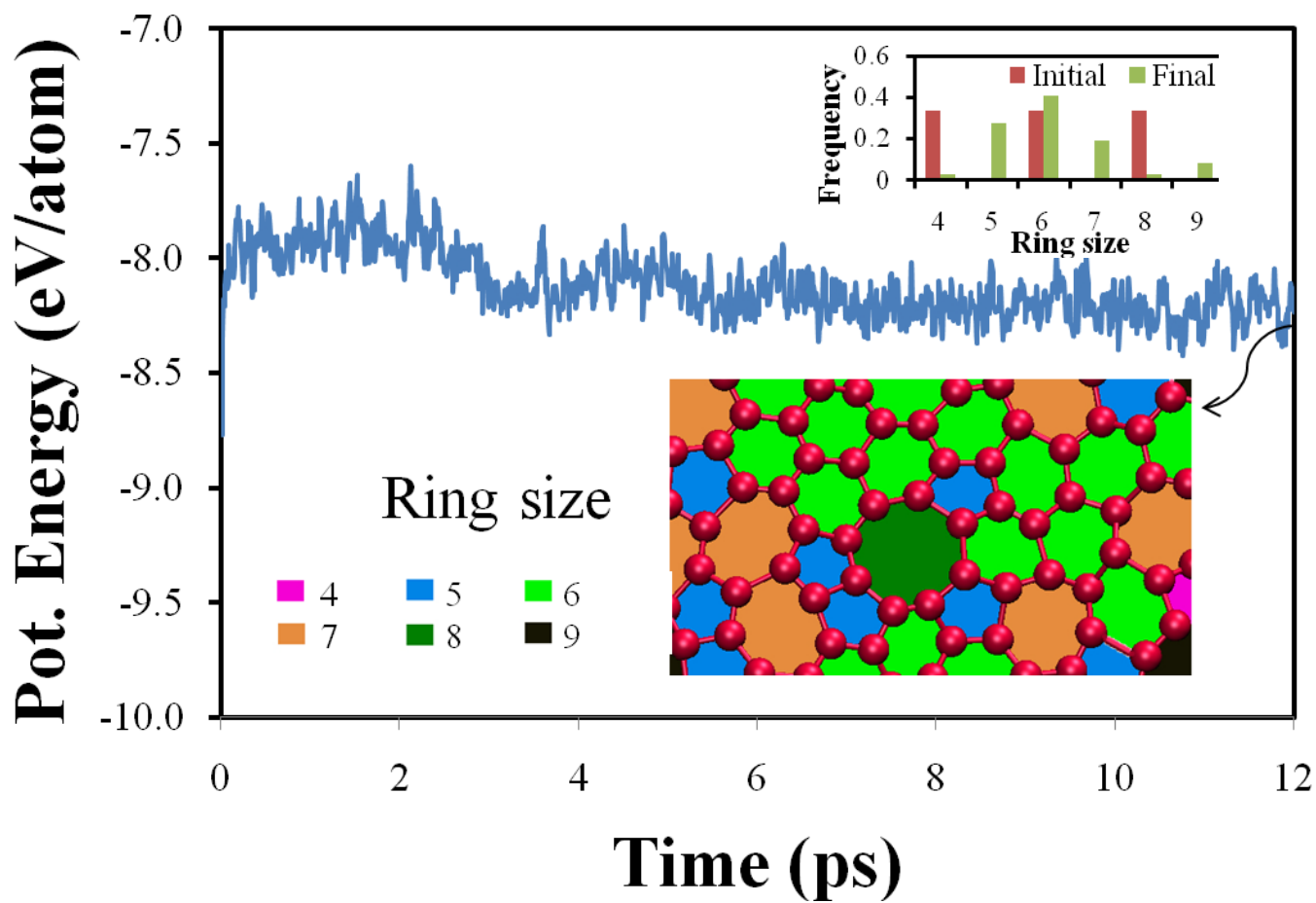

Figure 5: Transformation of biphenylene at 5000K. The final snapshot shows the rings of different sizes with different colors. The inset shows the distribution of ring sizes at the initial and final configuration.

When the final structure was minimized, its energy was found to be -9.02 eV/atom, only 0.26 eV/atom higher than the energy of hexa-graphene. Although it cannot be confirmed by this relatively short length of simulation (due to the high computational cost), a full transformation into pure hexa-graphene is possible in an elongated simulation.

This transformation of biphenylene into a hexa-graphene like structure demonstrates that although biphenylene is thermally stable, heating it beyond a threshold temperature can overcome the energy barrier into a more stable structure. Thus, thermal annealing and quenching can be exploited to synthesize one allotrope of 2D carbon from another. We note that an accurate estimation of this threshold temperature is not feasible within the framework of this work due to the limitation of simulation time.

It is important to note that the melting temperature of graphene is about 4510 K.[59] Thus, the retention of biphenylene structure at a temperature as high as 4000K is impressive considering that it is a less stable structure than graphene. On the other hand, the transformation of biphenylene into the graphene-like structure at 5000K can be simply regarded as melting since this temperature is higher than the melting temperature of graphene. If this temperature is sufficient to melt the most stable structure graphene it would possibly melt the less stable structure biphenylene as well. However, it is interesting to note that the melting of biphenylene goes through an intermediate that looks like graphene.

**Biphenylene is robust against defects**

Next we verified the robustness of biphenylene against point defects. A supercell of 72 atoms was used for the test. The selected defects included two cases of mono-vacancy by removing a 1) C1 atom and 2) C2 atoms, three cases of di-vacancies by removing 1) a pair of C1 atoms 2) a pair of C2 atoms and 3) one C1 and one C2 atoms, two cases of adatoms by adding a pair of C atoms across the 1) C1-C2 bonds and 2) C2-C2 bonds, and two cases of Stone-Wales-like defects by rotating 1) one C1 and one C2 atoms and 2) a pair of C2 atoms by 90° (Figure 6). It is important to note that with the supercell used in this study the defect concentration is about $4.9 \times 10^{13}$ cm$^{-2}$, far exceeding the experimentally observed value of $3 \times 10^{11}$ cm$^{-2}$ in graphene.[60]

By looking at the minimized structures, it can be inferred that the introduction of the defects did not change the global structure of biphenylene. Also the structure remained planar except the two cases of adatoms. The distortions are local by nature suggesting robustness and stability of the structure against defects.

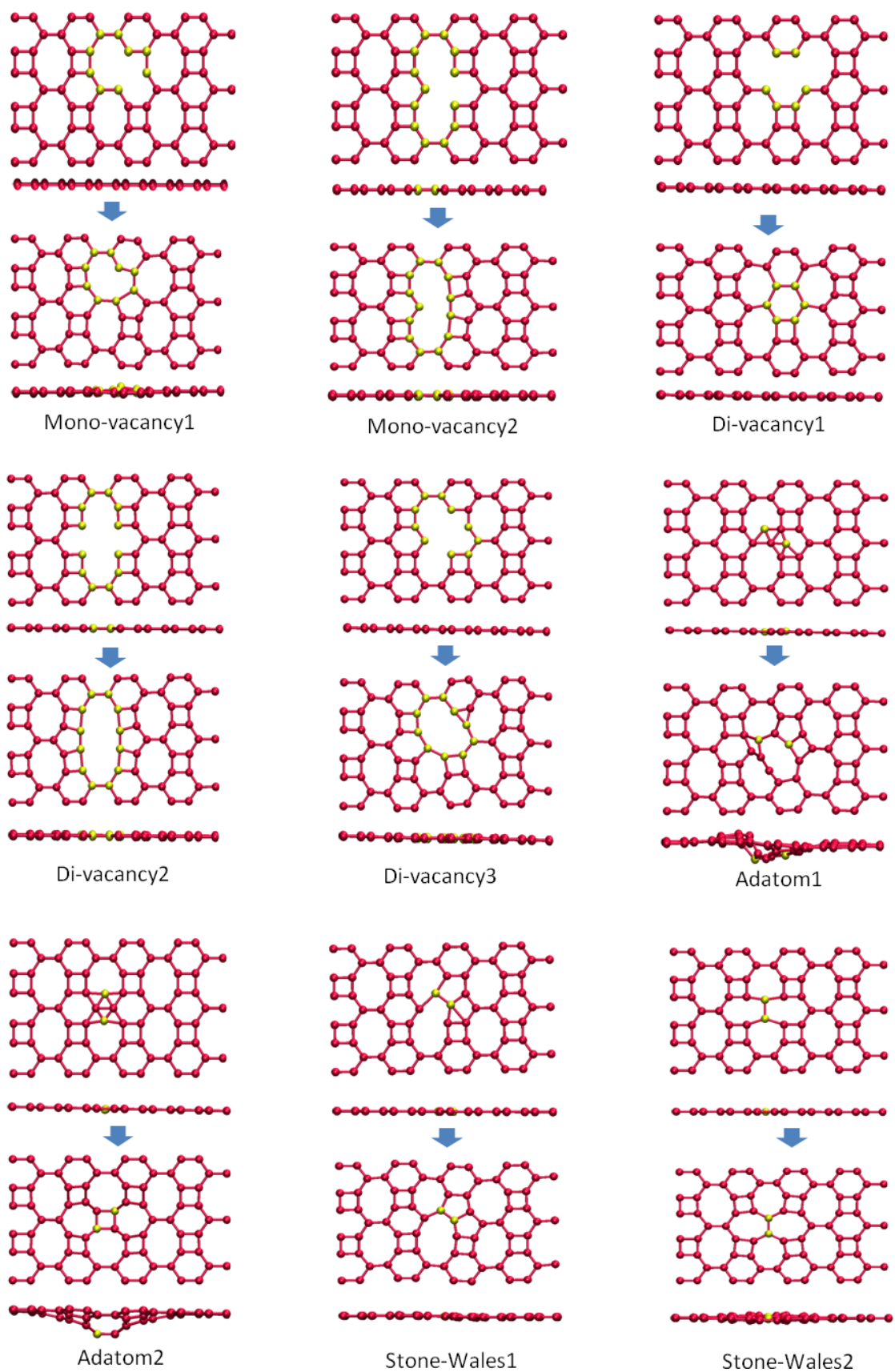

Mono-vacancy1     Mono-vacancy2     Di-vacancy1

Di-vacancy2     Di-vacancy3     Adatom1

Adatom2     Stone-Wales1     Stone-Wales2

Figure 6: Local defects in biphenylene. Both the initial and minimized structures are show. The atoms involved in the local distortions are shown in yellow.

In order to verify the thermal stabilities of the defected structures, AIMD simulations at T=300K were performed for 5 ps. The defected structures still remained intact at the end of the simulations (Figure S4). The potential energies of the structures also remained equilibrated.

Thus, it can be safely inferred that the point defects studied in this work affect biphenylene only locally, without disturbing its structural integrity and thermal stability.

**Biphenylene is mechanically strong**

The utility of biphenylene as a 2D carbon nano-material heavily depends on its mechanical strength and stability. In order to estimate the mechanical strength of biphenylene we calculated the change in stress under uniaxial strain (Figure 7). A unidirectional strain was applied while the box size in the transverse direction was adjusted so that the stress in that direction was close to zero. The simulation was repeated in both X and Y directions. A regime of elastic stretching under tensile loading was observed until about the strain value of 2%. Then the material went in the non-elastic region eventually reaching the breaking point. The breaking point was reached sooner in the X direction as compared to the Y direction. For the calculation of the stress we assumed the thickness of biphenylene as 3.35 Å, the same value as the thickness of graphene.

The estimated values of Young's modulus were $613 \pm 35$ GPa and $716 \pm 45$ GPa in the X and Y directions, respectively. This is less than but comparable to the Young's modulus of graphene (about 1000 GPa), one of the strongest material ever discovered.[61, 62] Thus, biphenylene demonstrates a notable mechanical strength making it reasonable to be used as a nanomaterial.

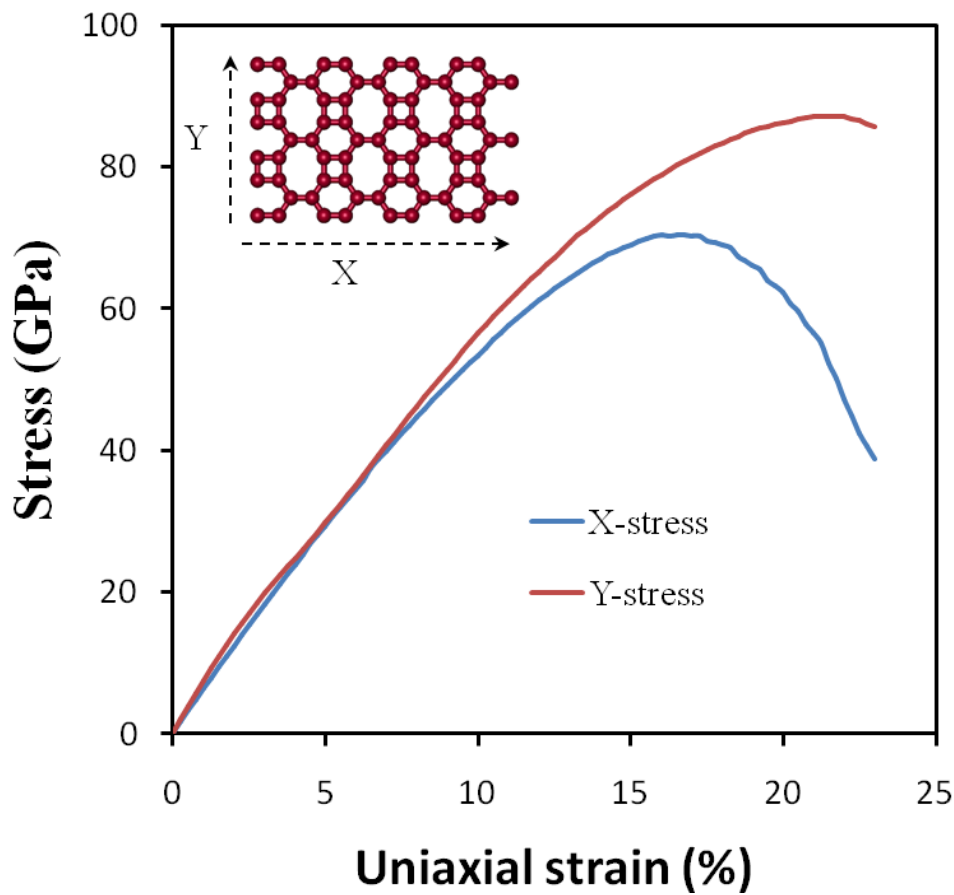

Figure 7: Biphenylene under uniaxial strain.

Our recent study on several graphene-like carbon and carbon nitride networks demonstrated proportionality between elastic modulus and atomic density (number of atoms divided by area).[51] Graphene which has a higher atomic density than biphenylene is expected to be stronger, which is confirmed by the results reported here.

**Biphenylene tube**

Next we examined the stability of biphenylene rolled up as a tube. A strip of biphenylene can be seamlessly rolled up to produce either an armchair shaped or a zigzag shaped tube depending on the orientations of the atoms along the tube's axis (Figure 8). We examined 3 armchair shaped (diameter= 4.44Å, 5.91Å and 7.27 Å) and 3 zigzag shaped (diameter= 4.94 Å, 6.01 Å and 8.40Å) tubes to determine their stabilities. All 6 structures, irrespective of their diameters or orientations, were intact after the energy minimization. All the tubular forms of biphenylene were found to be metallic in nature irrespective of their chirality and diameter (data not shown). This observation is in agreement with a previous DFT study on biphenylene[38] and other metallic 2-D allotropes of Carbon.[11] For a similar length of the diameter, the energy of the zigzag structure was found to be slightly lower than that of the armchair structure (Figure S5). The energy/atom decreased with increasing length of the diameter. This is expected as energy is released with the relaxation of the strain due to bending.

Bending stiffness of a 2D material is an important measure of its stability and strength. Bending stiffness can be calculated from the energy curvature data by fitting the following equation: $U_{bend}$ = 1/2 D $\kappa^2$, where $U_{bend}$ is the system strain energy divided by the basal plane area, D is the bending stiffness and $\kappa$ is the curvature of the tube.[63] The estimated bending stiffness for the zigzag and armchair structures were 1.86 eV and 1.96 eV, respectively. These values are within the range reported for monolayer graphene, i.e. from 0.69 eV to 2.1 eV.[63]

Next we studied the thermal stabilities of the six tubes by running six independent molecular dynamics simulations at 300K. After 5ps of simulations, all the structures were found to be stable at room temperature. The final snapshots and the evolutions of potential energies with time are shown in Figure S6. Again the zigzag structure was found to be slightly more stable than the armchair shaped structure for a similar length of the diameter.

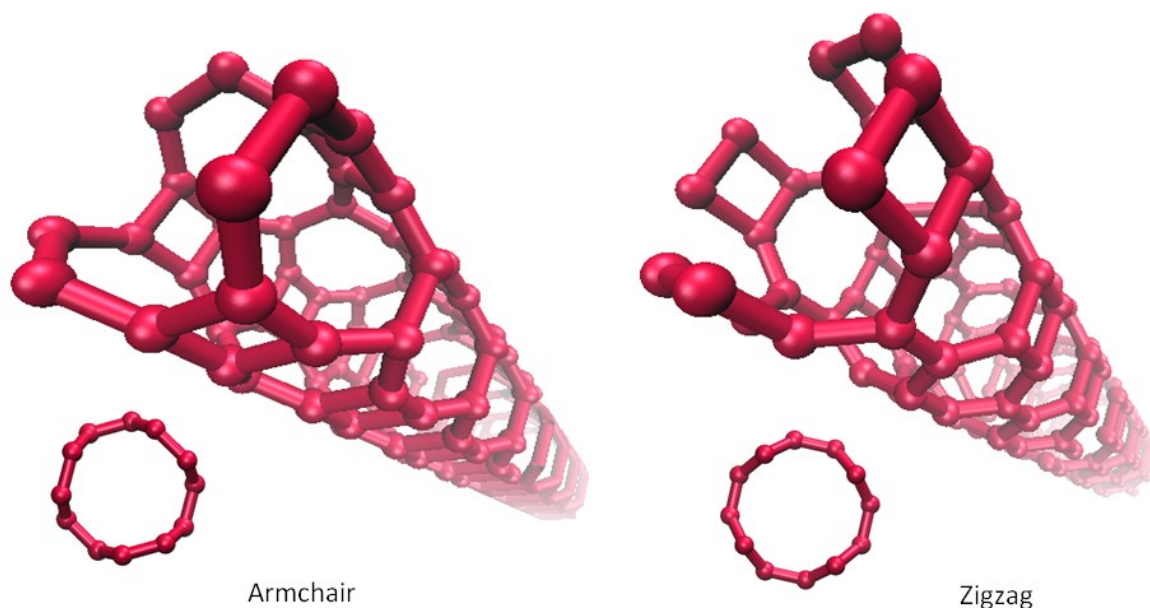

Figure 8: Armchair shaped (diameter= 4.44Å) and Zigzag shaped (diameter= 4.94 Å) biphenylene tubes,

**Conclusion**

We used density functional theory to demonstrate a sudden phase transformation of penta-graphene into biphenylene under uniaxial tension. This was accompanied by a notable drop in the energy indicating a meta-stable nature of penta-graphene. Our first principles study highlights that the application of uniaxial tension can be considered as one of the methods to verify the stability of a proposed material.

This was followed by a thorough characterization of biphenylene. Our investigation suggests that biphenylene is energetically, mechanically, dynamically and thermally a stable 2-D carbon allotrope. It shows metallic nature both in the form of a sheet and tube. We also demonstrated that biphenylene is robust against defects. Another phase transformation from biphenylene to a hexa-graphene like structure was observed at a high temperature of 5000K. This suggests that thermal annealing can be exploited as a novel method of synthesizing one 2-D carbon allotrope from another.

This study not only confirms that penta-graphene is a meta-stable form of graphene but also demonstrates the possibility of phase transformation in 2D materials using mechanical deformation. This is a step toward understanding phase transition in 2D materials at a fundamental level. This also opens up a new paradigm in 2D materials engineering where the properties of a material can be manipulated by regulating its phase. For instance, this work demonstrates the transformation of a semiconductor (penta-graphene) into a conductor (biphenylene) by applying uniaxial tension. Although penta-graphene is not yet realized experimentally, with rapid growth in technology its synthesis might be possible in the future. In light of the insights obtained in this work, penta-graphene, as well as any other meta-stable form of carbon, if synthesized, opens up the possibility of novel technological applications based on regulated phase transition.

**Acknowledgements**: The authors O. R., B. M. and T. R. gratefully acknowledge the financial support of the European Research Council (Grant number 615132).

**Supplementary Information**

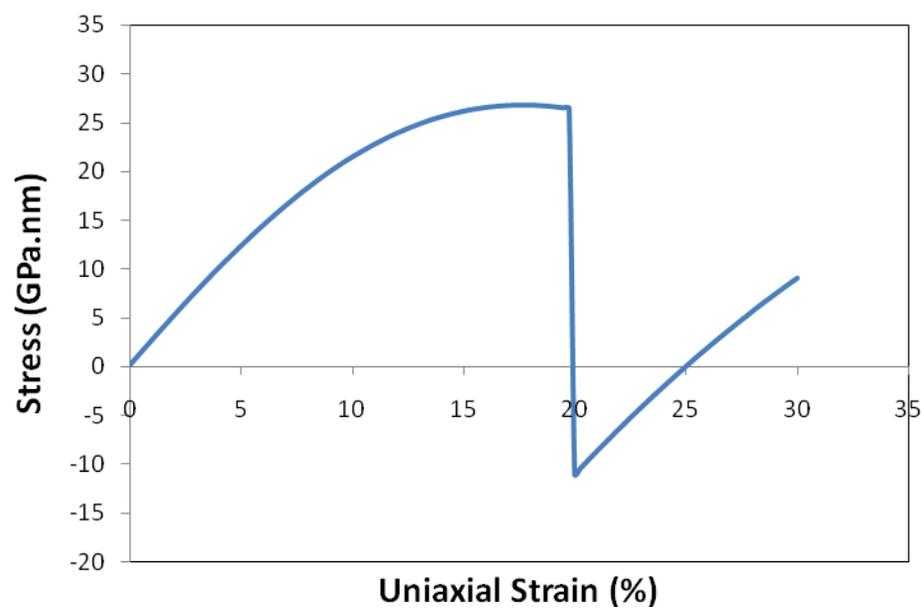

Figure S1: Stress-strain curve of penta-graphene to biphenylene transition

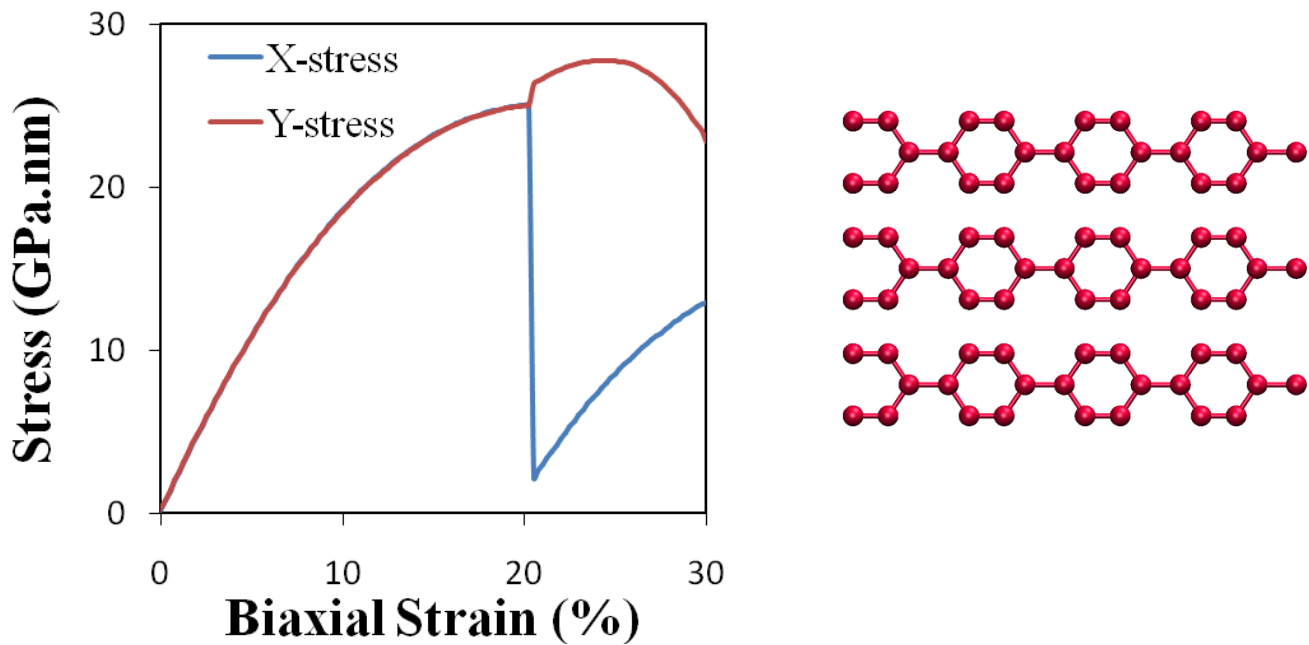

Figure S2: Stress-strain curve of penta-graphene under biaxial strain and the final product.

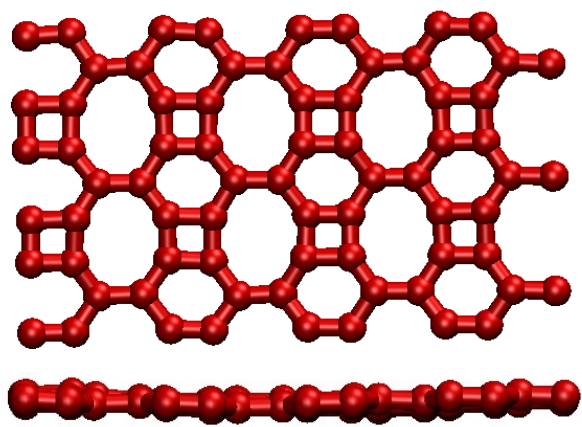

T = 300K, t=6ps

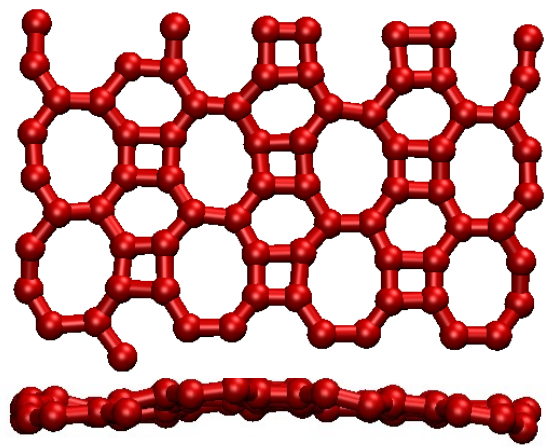

T = 1000K, t=6ps

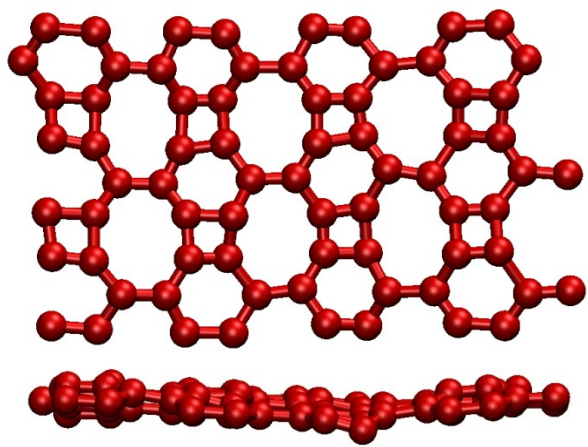

T = 2000K, t=6ps

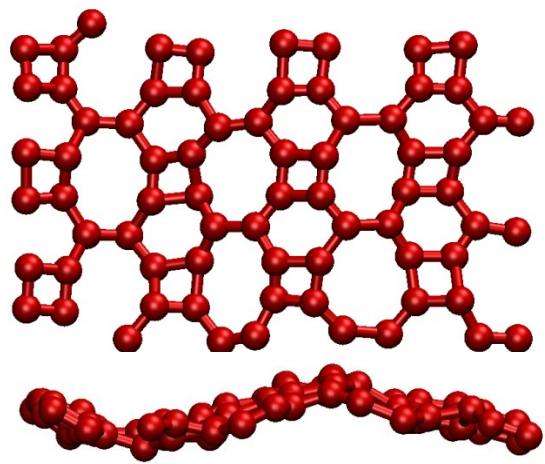

T = 3000K, t=6ps

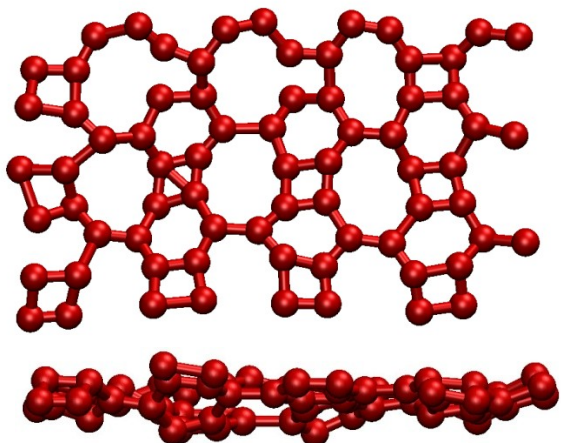

T = 4000K, t=12ps

Figure S3: The final snapshots of AIMD simulation of biphenylene at different temperatures and simulation times.

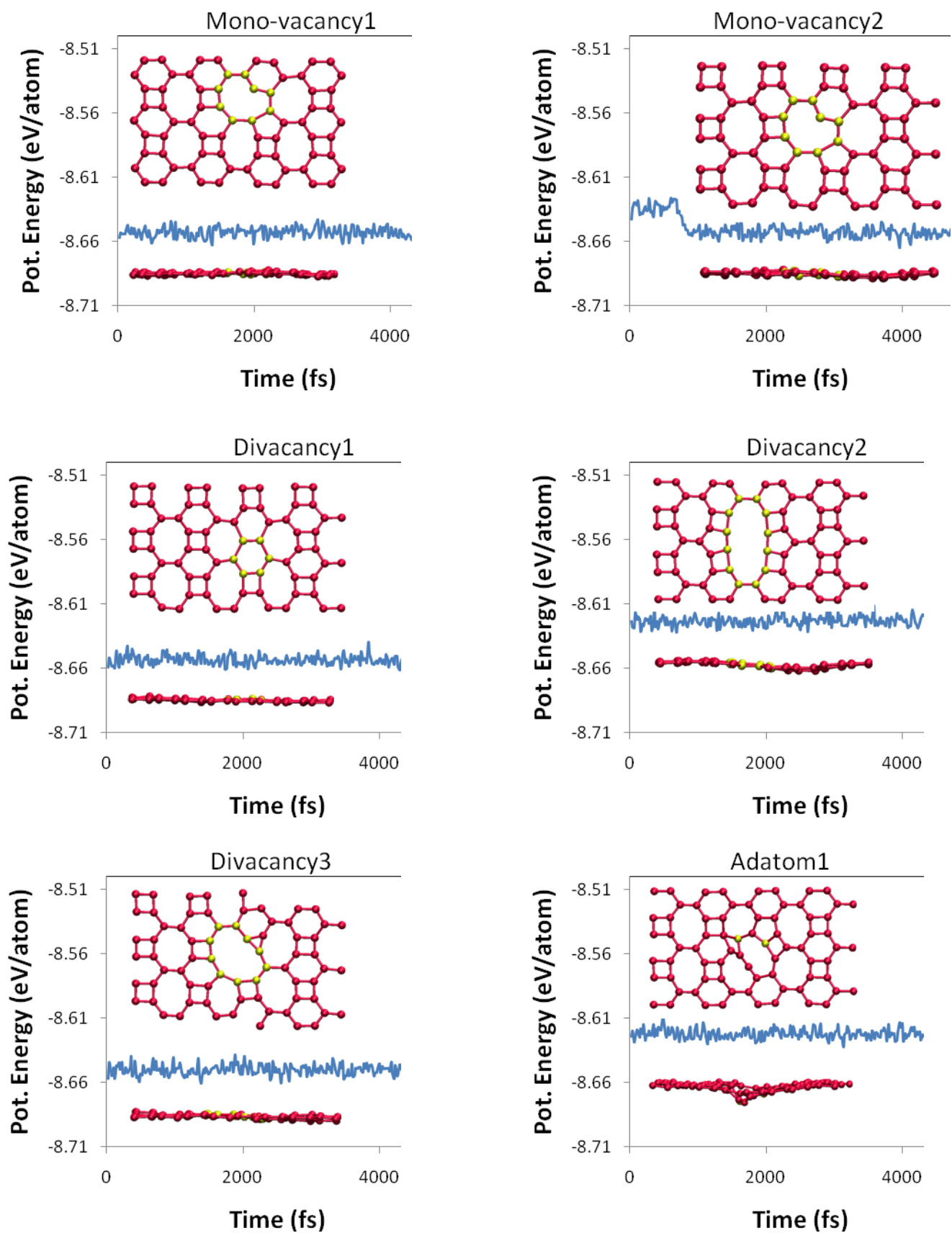

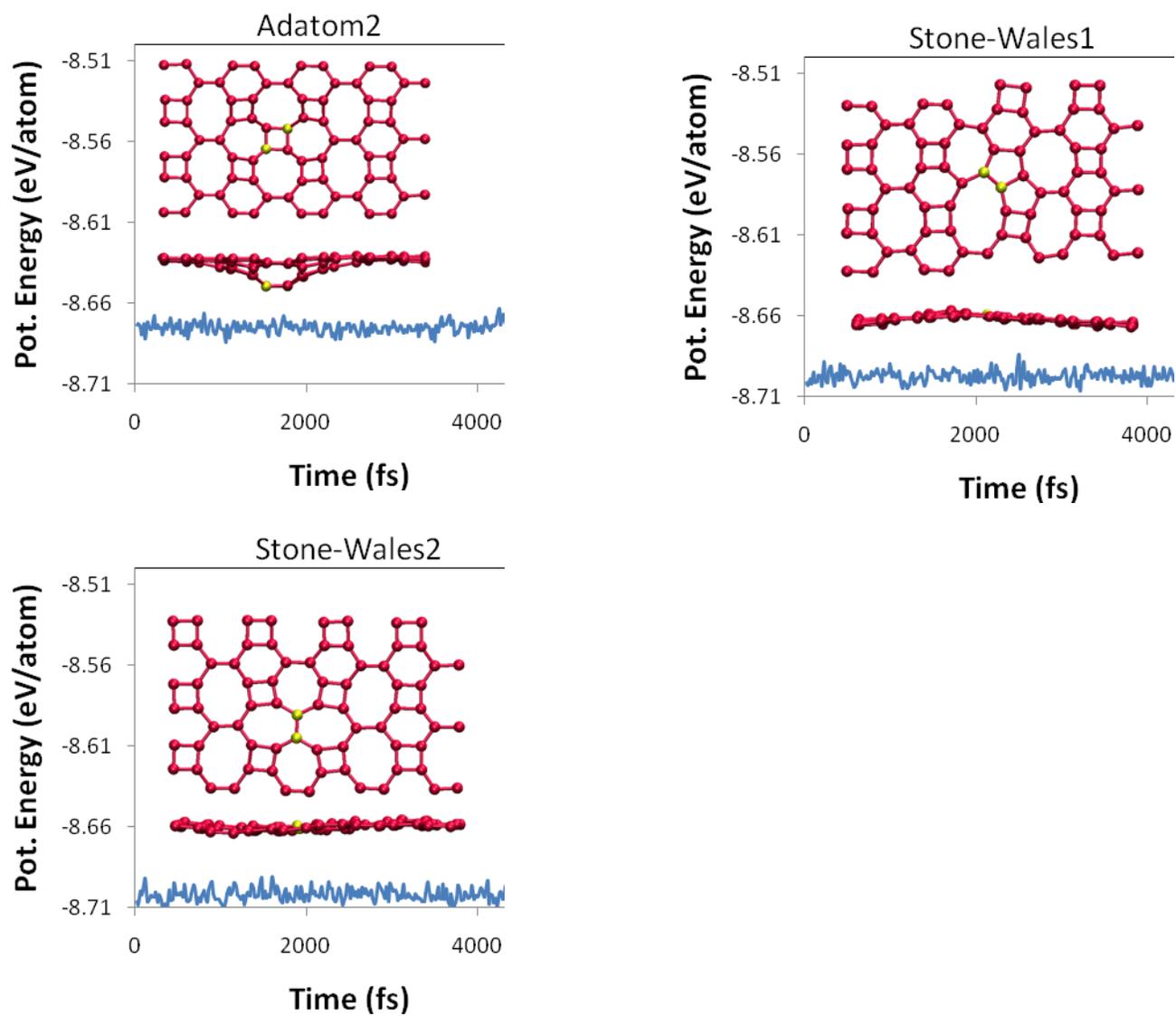

Figure S4: The final snapshots of AIMD simulation of biphenylene with defects at T=300K

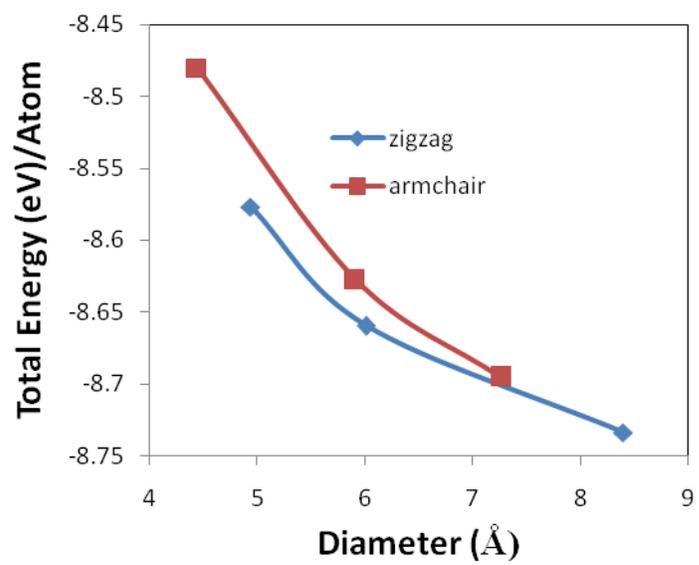

Figure S5: Comparison of the energies of biphenylene nanotubes of different diameters.

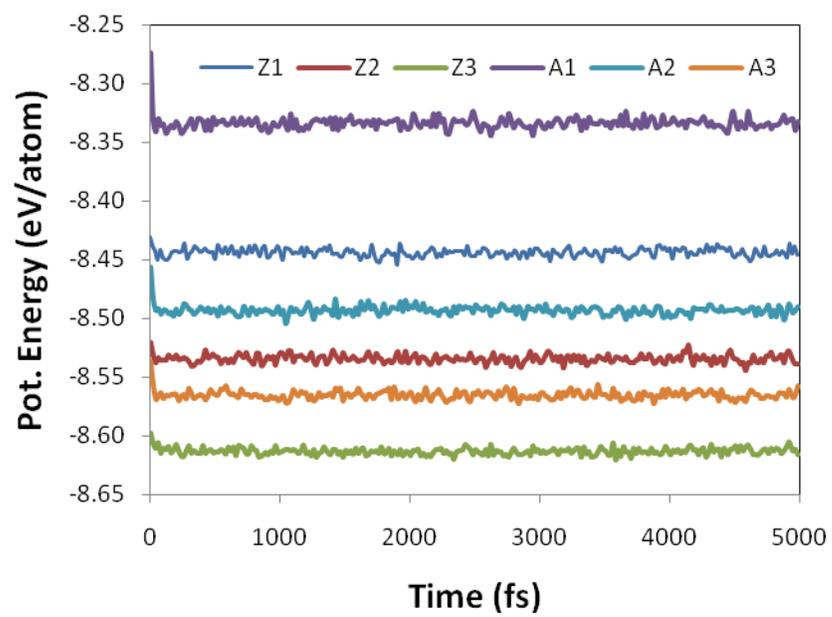

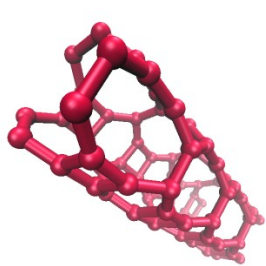

A1

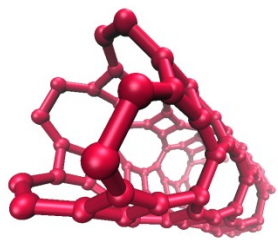

A2

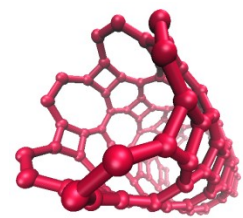

A3

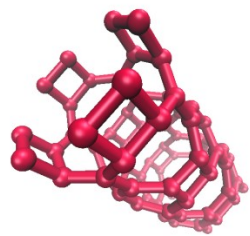

Z1

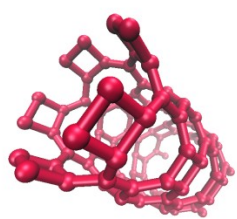

Z2

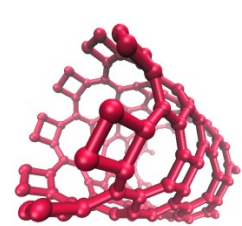

Z3

Figure S6: Potential energies and final snapshots of three armchair shaped (A1, A2 and A3) and three zigzag shaped (Z1, Z2 and Z3) tubes of biphenylene. The molecular dynamics simulations were performed at T=300K